\documentclass[a4paper,3p,twocolumn,preprint]{elsarticle}

 \journal{Wave Motion}

\usepackage{amsmath}
\usepackage{amssymb}
\usepackage{graphicx}% Include figure files
\usepackage{bm}% bold math
\usepackage{epstopdf}

\newcommand{\ga}{\gamma}

\newcommand{\om}{\omega}

\newcommand{\prt}{\partial}

\begin{document}

\begin{frontmatter}

\title{
Propagation of dark solitons of DNLS equation along a large-scale
background}

\author{A.~M.~Kamchatnov}
\ead{kamch@isan.troitsk.ru}
\author{D.~V.~Shaykin}
\ead{shaykin.dv@phystech.edu}
\address{
Institute of Spectroscopy, Russian Academy of Sciences, Troitsk,
Moscow, 108840, Russia }
\address{Moscow Institute of Physics and Technology, Institutsky lane 9, Dolgoprudny, Moscow region, 141700, Russia}
\address{
Skolkovo Institute of Science and Technology, Skolkovo, Moscow, 143026, Russia}

%\date{\today}

\begin{abstract}
We study dynamics of dark solitons in the theory of the DNLS equation by the method
based on imposing the condition that this dynamics must be Hamiltonian. Combining
this condition with Stokes' remark that relationships for harmonic linear waves and
small-amplitude soliton tails satisfy the same linearized equations, so the corresponding
solutions can be converted one into the other by replacement of the packet's wave
number $k$ by $i\kappa$, $\kappa$ being the soliton's inverse half-width, we find
the Hamiltonian and the canonical momentum of the soliton's motion. The Hamilton
equations are reduced to the Newton equation whose solutions for some typical situations
are compared with exact numerical solutions of the DNLS equation.
\end{abstract}


%02.30.Ik Integrable systems
%05.45.Yv Solitons
%43.20.Bi Mathematical theory of wave propagation

\end{frontmatter}

%\maketitle

{\it Dedicated to the memory of Noel Smyth}

\section{Introduction}

In situations when the soliton's width is much smaller than the typical length of the
background wave along which the soliton propagates, one can introduce with good enough
accuracy the soliton's coordinate $x(t)$ and describe its propagation as a motion of a 
point-like particle through a non-uniform and varying with time surrounding. In this
case, evolution of the background wave is governed by the equations of dispersionless
(hydrodynamic) approximation independently of the soliton's motion. However, the soliton's
motion cannot be separated from the background wave evolution: this motion causes a 
counterflow around the soliton and such a counterflow changes drastically the soliton's
dynamics. Well-known examples of this back reaction on the soliton's motion are the
formation of shelves behind KdV solitons propagating along shallow water with uneven 
bottom (see, e.g., \cite{km-77,km-78,kn-80,newell-85}) and change of the frequency of
oscillations of a dark soliton in a Bose-Einstein condensate confined in a harmonic
trap (see, e.g., \cite{ba-2000,kp-04}).

So far, the counterflow effects were studied by different forms of perturbation analysis.
It was recently noticed \cite{ks-23a}, that if the nonlinear wave equation under consideration
can be written in a Hamiltonian form and one assumes that the reduction of the wave evolution
to the soliton's motion along the large-scale background wave remains Hamiltonian, then such 
an analysis can
be considerably simplified. In fact, it is well known that completely integrable equations
are Hamiltonian (see, e.g., \cite{zf-71,gardner-71,dickey-03}), but their Hamiltonian dynamics 
is formulated in terms of the scattering data or related variables in the inverse scattering
transform method. Since a large-scale background wave can be represented as a collection of
a large number of long waves associated with the continuous spectrum of the linear spectral problem,
the assumption that the soliton's particle-like dynamics remains Hamiltonian seems very
natural. This idea was recently applied to dynamics of KdV \cite{ks-23a} and NLS \cite{kamch-24}
solitons where it was also generalized to some non-integrable situations. In this paper, our
goal is to apply this method to propagation of the DNLS soliton.

The DNLS equation
\begin{equation}\label{eq1}
  i\psi_t+\frac12\psi_{xx}-i(|\psi|^2\psi)_x=0
\end{equation}
has applications to propagation of the nonlinear Alfv\'{e}n waves in magnetized plasma
(see, e.g., \cite{kbhp-88} and references therein) and of ultrashort optical pulses in nonlinear
fibers (see, e.g., \cite{avc-86}). Complete integrability of this equation was established in
Ref.~\cite{kn-78}. We will not use here this fact explicitly (except for the assumption
that the soliton's dynamics is Hamiltonian) and we will obtain first the Hamilton equations
which govern the soliton's motion along a non-uniform and time-dependent background wave.
Then we derive the Newton equation which is more convenient for applications and, at last,
compare our analytical theory with numerical solutions of Eq.~(\ref{eq1}) for the initial 
conditions corresponding to propagation of solitons along the large-scale rarefaction waves or
non-uniform profiles created by an external potential.

\section{Basic equations}

Substitution 
\begin{equation}\label{eq2}
  \psi(x,t)=\sqrt{\rho(x,t)}\exp\left(i\int^xu(x',t)d x'\right)
\end{equation}
and separation of real and imaginary parts transform Eq.~(\ref{eq1}) to the hydrodynamic-like form
\begin{equation}\label{eq3}
\begin{split}
   &\rho_t+\left[\rho\left(u-\frac32\rho\right)\right]_x=0,\\
   &u_t+uu_x-(\rho u)_x+\left(\frac{\rho_x^2}{8\rho^2}-\frac{\rho_{xx}}{4\rho}\right)_x=0,
   \end{split}
\end{equation}
for real variables $\rho$ and $u$ which can be interpreted as ``density'' and ``flow velocity'',
correspondingly. This system has two important limits.

First, we can consider small-amplitude waves propagating along a uniform background with
constant values of $\rho=\rho_0$ and $u=u_0$. Then linearization of Eqs.~(\ref{eq2}) with
respect to small deviations $\rho'=\rho-\rho_0$, $u'=u-u_0$ leads after standard calculations
to the dispersion relation for the harmonic waves $\rho',u'\propto \exp[i(kx-\om t)]$:
\begin{equation}\label{eq4}
  \omega(k)=k\left[u_0-2\rho_0\pm\sqrt{\rho_0(\rho_0-u_0)+k^2/4}\right].
\end{equation}
As we see, the uniform state $(\rho_0,u_0)$ becomes modulationally unstable for $\rho_0<u_0$,
so we will only confine ourselves to the modulationally stable situations with $\rho_0>u_0$.

Second, we can consider large-scale waves with typical wavelength much greater than the
dispersion size $O(1)$ in our non-dimensional variables, $|\rho_x/\rho|,|u_x/u|\ll1$.
Then we can neglect the dispersive term in Eqs.~(\ref{eq3}) and arrive at the dispersionless
equations
\begin{equation}\label{eq5}
    \rho_t+\left[\rho\left(u-\frac32\rho\right)\right]_x=0,\quad
  u_t+uu_x-(\rho u)_x=0.
\end{equation}
The characteristic velocities of this system are equal to
\begin{equation}\label{eq6}
  v_{\pm}=u-2\rho\pm\sqrt{\rho(\rho-u)}.
\end{equation}
Naturally, they are real for $\rho>u$ and coincide with the phase velocities 
$\left.\omega/k\right|_{k\to0}$ of linear waves with the dispersion relation (\ref{eq4})
in the long wavelength limit. As follows from the condition $u<\rho$, they both are
negative, $v_-<v_+<0$. Eqs.~(\ref{eq5}) can be cast to the Riemann diagonal form
\begin{equation}\label{eq7}
\frac{\prt r_{\pm}}{\prt t}+v_{\pm}\frac{\prt r_{\pm}}{\prt x}=0.
\end{equation}
for the Riemann invariants
\begin{equation}\label{eq8}
  r_{\pm}=\frac{u}2-\rho\pm\sqrt{\rho(\rho-u)},
\end{equation}
and the velocities (\ref{eq6}) are expressed in terms of Riemann invariants by the formulas 
\begin{equation}\label{eq9}
  v_+=\frac32r_++\frac12r_-,\qquad v_-=\frac12r_++\frac32r_-.
\end{equation}
If some solution of Eqs.~(\ref{eq7}) is found, then the physical variables are given 
by the expressions 
\begin{equation}\label{eq10}
  \rho=\frac12(\sqrt{-r_+}\pm\sqrt{-r_-})^2,\qquad u=\pm 2\sqrt{r_+r_-}.
\end{equation}
These equations describe evolution of the background wave.

\section{DNLS soliton}

Soliton solutions of the DNLS equation (\ref{eq1}) were found in Refs.~\cite{kbhp-88,kn-78,kamch-90}.
Here we reproduce this solution in a convenient for us form.

If we look for the traveling wave solution of Eqs.~(\ref{eq3}) in the form $\rho=\rho(\xi)$,
$u=u(\xi)$, $\xi=x-Vt$, $V$ being the phase velocity of the wave, then Eqs.~(\ref{eq3}) becomes
a pair of ordinary differential equations which can be easily integrated to give
\begin{equation}\label{eq11}
  u(\xi)=\frac{A}{\rho(\xi)}+V+\frac32\rho.
\end{equation}
\begin{equation}\label{eq12}
\begin{split}
  \rho_{\xi}^2(\xi)=&-\mathcal{R}(\rho)=-\big[\rho^4+4V\rho^3\\
  &+4(V^2-A+B)\rho^2-4C\rho+4A^2\big],
  \end{split}
\end{equation}
where $A,B,C$ are the integration constants. The solution is parameterized more conveniently
by the zeroes $\nu_i$, $i=1,2,3,4,$ of the polynomial $\mathcal{R}(\rho)$,
\begin{equation}\label{eq13}
  \mathcal{R}(\rho)=\prod_{i=1}^4
(\rho-\nu_i)=\rho^4-s_1\rho^3+s_2\rho^2-s_3\rho+s_4,
\end{equation}
where
\begin{equation}\label{eq14}
  \begin{split}
  & s_1=\sum_i\nu_i=-4V,\\
  & s_2=\sum_{i<j}\nu_i\nu_j=4(V^2-A+B),\\
  & s_3=\sum_{i<j<k}\nu_i\nu_j\nu_k=4C,\\
  & s_4=\nu_1\nu_2\nu_3\nu_4=4A^2.
  \end{split}
\end{equation}
We assume that $\nu_i$ are ordered according to
\begin{equation}\label{eq15}
  \nu_1\leq\nu_2\leq\nu_3\leq\nu_4.
\end{equation}

The soliton solutions appear when two middle zeroes coincide, $\nu_2=\nu_3=\rho_0$, and then they
are equal to the background density $\rho_0$ of the uniform state along which the soliton
propagates. If $\rho$ varies within the interval $\nu_1\leq\rho\leq\nu_2=\rho_0$, then we obtain a
dark soliton; if within the interval $\nu_2=\rho_0\leq\rho\leq\nu_4$, then the soliton is bright.
To be definite, we confine ourselves to the case of dark solitons with $\nu_1\leq\rho\leq\nu_2=\rho_0$.
The standard calculation yields the solution of Eq.~(\ref{eq12}) in  the form
\begin{equation}\label{eq16}
  \rho=\rho_0-\frac{(\rho_0-\nu_1)(\nu_4-\rho_0)}{(\nu_4-\nu_1)\cosh^2\theta-(\rho_0-\nu_1)},
\end{equation}
where $\theta=\frac12\kappa\xi$ and
\begin{equation}\label{eq17}
  \kappa=\sqrt{(\rho_0-\nu_1)(\nu_4-\rho_0)}
\end{equation}
is the inverse half-width of the soliton. 

Let us express $\nu_1,\nu_4$ in terms of the values
$(\rho_0,u_0)$ of the parameters of the flow at infinity. Eq.~(\ref{eq11}) and the first
Eq.~(\ref{eq14}) give
$$
\nu_1+\nu_4=-4V-2\rho_0,\quad \nu_1\nu_4=(2u_0-3\rho_0-2V)^2,
$$
so we get
\begin{equation}\label{eq18}
  \begin{split}
  & \nu_1=-2V-\rho_0-\sqrt{(\rho_0-u_0)(u_0-2\rho_0-2V)},\\
  & \nu_4=-2V-\rho_0+\sqrt{(\rho_0-u_0)(u_0-2\rho_0-2V)}.
  \end{split}
\end{equation}
Substitution of these values into Eq.~(\ref{eq17}) yields
\begin{equation}\label{eq19}
  \kappa^2=4\left[\rho_0(\rho_0-u_0)-(V+2\rho_0-u_0)^2\right].
\end{equation}
Whence, the soliton velocity $V$ is related with the inverse half-width $\kappa$ by the formula
\begin{equation}\label{eq20}
  V=u_0-2\rho_0\pm\sqrt{\rho_0(\rho_0-u_0)-\frac{\kappa^2}{4}}.
\end{equation}
Comparison with Eq.~(\ref{eq4}) shows that the soliton's velocity is related with the dispersion
law of linear waves by the expression
\begin{equation}\label{eq21}
  V=\frac{\om(i\kappa)}{i\kappa}.
\end{equation}
This is a particular case of the general Stokes' remark \cite{stokes} (see also \cite{lamb,kamch-20})
that both linear waves $\propto\exp[i(kx-\om t)]$ and soliton's tails $\propto\exp[\pm\kappa(x-Vt)]$
are described by the linearized equations, so the phase velocity $\om(k)/k$ transforms to the
soliton's velocity by the replacement $k\to i\kappa$.

\begin{figure}[t]
\begin{center}
	\includegraphics[width = 7cm,height = 5cm]{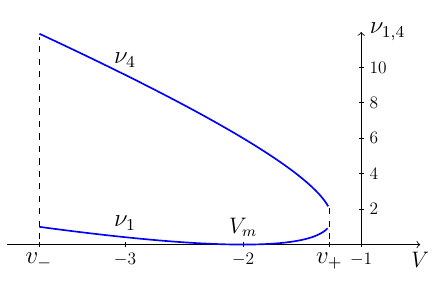}
\caption{Dependence of $\nu_1$ and $\nu_4$ on the soliton's velocity $V$ ($\rho_0=1,u_0=-0.5$).
 }
\label{fig1}
\end{center}
\end{figure}

\begin{figure}[t]
\begin{center}
	\includegraphics[width = 7cm,height = 5cm]{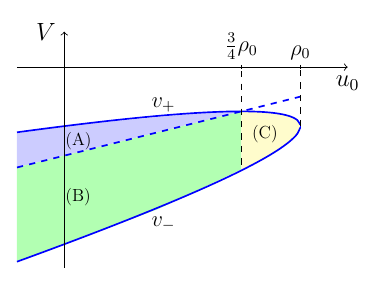}
\caption{Regions of the soliton's velocity $V$ at different values of $u_0$ and fixed $\rho_0$
with different expressions for soliton's phase. They are confined between the boundaries
(\ref{eq23}) and separated by the lines $V_m(u_0)=u_0-\frac32\rho_0$ and $u_0=\frac34\rho_0$.
 }
\label{fig2}
\end{center}
\end{figure}

If we rewrite Eq.~(\ref{eq19}) in the form 
\begin{equation}\label{eq22}
\begin{split}
  \kappa^2&=4\left[\sqrt{\rho_0(\rho_0-u_0)}-2\rho_0+u_0-V\right]\\
  &\,\,\,\,\,\times\left[\sqrt{\rho_0(\rho_0-u_0)}+2\rho_0-u_0+V\right],
  \end{split}
\end{equation}
we find at once that the soliton's velocity $V$ must have the value between the dispersionless
characteristic velocities (\ref{eq5}),
\begin{equation}\label{eq23}
  v_-<V<v_+,
\end{equation}
that is it is always negative. The parameters $\nu_1$ and $\nu_4$ are always positive since
positive density at the center of soliton is equal to $\rho_{min}=\nu_1$ and $\nu_4>\nu_1$.
One can easily find that at the edges of the interval (\ref{eq23}) the parameters $\nu_1$
and $\nu_4$ are equal to
\begin{equation}\label{eq24}
  \begin{split}
  &\nu_1(v_-)=\rho_0,\\
  &\nu_1(v_+)=\left\{
                \begin{array}{ll}
                  \rho_0, & \hbox{for}\quad u_0<0, \\
                  5\rho_0-4u_0\\-4\sqrt{\rho_0(\rho_0-u_0)}, & \hbox{for}\quad 0<u_0<\rho_0;
                \end{array}
              \right.
  \end{split}
\end{equation}
and
\begin{equation}\label{eq25}
  \begin{split}
  &\nu_4(v_-)=5\rho_0-4u_0+4\sqrt{\rho_0(\rho_0-u_0)},\\
  &\nu_4(v_+)=\left\{
                \begin{array}{ll}
                  5\rho_0-4u_0\\-4\sqrt{\rho_0(\rho_0-u_0)}, & \hbox{for}\quad u_0<0, \\
                  \rho_0, & \hbox{for}\quad 0<u_0<\rho_0.
                \end{array}
              \right.
  \end{split}
\end{equation}
Plots of $\nu_1(V)$ and $\nu_4(V)$ are shown in Fig.~\ref{fig1}; $\nu_1(V)$ has vanishing minimal value at
\begin{equation}\label{eq26}
  V=V_m=u_0-\frac32\rho_0.
\end{equation}
One can easily find that this point is located inside the interval (\ref{eq23}), 
that is $V_m\leq v_+(u_0)$, for $u_0< u_m=\frac34\rho_0$, and $v_+(u_0)$ has its maximal
value $v_+(u_m)=-\frac18\rho_0$ at this point. Thus, for $u_0<\frac34\rho_0$ we should
distinguish two intervals of the soliton's velocity:
\begin{equation}\label{eq27}
  \begin{split}
\mathrm{(A)}: \quad u_0-\frac32\rho_0<V<v_+,\quad V+\frac32\rho_0-u_0>0,\\
\mathrm{(B)}: \quad v_-<V<u_0-\frac32\rho_0,\quad V+\frac32\rho_0-u_0<0.
\end{split}
\end{equation}
If $\frac34\rho_0<u_0<\rho_0$, then we get a single interval
\begin{equation}\label{eq28}
  \mathrm{(C)}:\quad v_-<V<v_+,\quad  V+\frac32\rho_0-u_0<0.
\end{equation}
These different intervals of the soliton's velocity are depicted in Fig.~\ref{fig2}; such their
separation is important for calculation of the soliton's phase.

\begin{figure}[t]
\begin{center}
	\includegraphics[width = 7cm,height = 5cm]{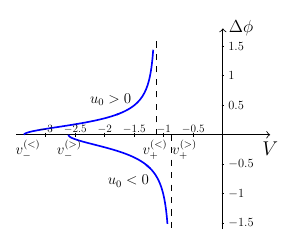}
\caption{Jump of phase across the soliton as a function of its velocity $V$.
It equals to $\Delta\phi(v_+)=\mathrm{sgn}(u_0)2\pi$ at the edge $V=v_+$. 
 }
\label{fig3}
\end{center}
\end{figure}

The soliton's phase in the solution $\psi=\sqrt{\rho}\exp(i\phi)$ can be found by integration 
of Eq.~(\ref{eq11}), that is
\begin{equation}\label{eq29}
  \phi=\int\left(\frac{\rho_0\sqrt{\nu_1\nu_4}}{\rho}+V+\frac32\rho\right)d\xi,
\end{equation}
where $\rho=\rho(\xi)$ is given in Eq.~(\ref{eq16}). Quite a tedious, although straightforward,
calculation yields
\begin{equation}\label{eq30}
  \begin{split}
\phi_A=u_0\xi&-\arctan\left(\frac{1}{\ga}\sqrt{\frac{\nu_4}{\nu_1}}\tanh\frac{\kappa\xi}{2}\right)\\
&-3\arctan(\left(\frac{1}{\ga}\tanh\frac{\kappa\xi}{2}\right),\\
\phi_{B,C}=u_0\xi&+\arctan\left(\frac{1}{\ga}\sqrt{\frac{\nu_4}{\nu_1}}\tanh\frac{\kappa\xi}{2}\right)\\
&-3\arctan(\left(\frac{1}{\ga}\tanh\frac{\kappa\xi}{2}\right),
\end{split}
\end{equation}
where
\begin{equation}\label{eq31}
  \ga=\sqrt{\frac{\nu_4-\rho_0}{\rho_0-\nu_1}}.
\end{equation}
Obviously, $u_0\xi$ is the background phase over which the soliton's phase is imposed. The jump of the phase
across the soliton is equal to
\begin{equation}\label{eq32}
  \begin{split}
&\Delta\phi_A=-2\arctan\left(\frac{1}{\ga}\sqrt{\frac{\nu_4}{\nu_1}}\right)-6\arctan\frac{1}{\ga}+2\pi,\\
&\Delta\phi_{B,C}=2\arctan\left(\frac{1}{\ga}\sqrt{\frac{\nu_4}{\nu_1}}\right)-6\arctan\frac{1}{\ga}.
\end{split}
\end{equation}
Its dependence on the soliton's velocity $V$ is shown in Fig.~\ref{fig3} for positive and negative 
signs of the background velocity $u_0$. At $V=v_+$ we get $\Delta\phi=\mathrm{sgn}(u_0)2\pi$ for all $u_0$,
but $\Delta\phi\sim|u_0|$, $u_0\to0$, tends to zero non-uniformly almost everywhere except a 
decreasing interval $v_+-|u_0|<V\leq v_+$
close to the edge $V=v_+$. So one can consider $\Delta\phi=0$ at $u_0=0$ for all $V<v_+$.

Combining expressions (\ref{eq16}) for $\rho$ and (\ref{eq30}) for $\phi$, we arrive at the soliton solution
\begin{equation}\label{eq33}
  \begin{split}
\psi_{A,B,C}=&\frac{\sqrt{\nu_4}\tanh(\kappa\xi/2)\pm i\sqrt{\nu_1}\ga}{[\ga+i\tanh(\kappa\xi/2)]^3}\\
&\times\left(\ga^2+\tanh^2\frac{\kappa\xi}{2}\right)e^{iu_0\xi},
\end{split}
\end{equation}
where ``+'' corresponds to the region A and ``--'' to the regions B,C. We will use this expression in
section~\ref{sec5} as an initial condition in numerical solutions of Eq.~(\ref{eq1}).

\section{Hamilton equations for soliton motion}

We assume that soliton's width is much smaller than the characteristic background wavelength.
Therefore, we can calculate the soliton's Hamiltonian $H$ and the canonical momentum $p$ under
assumption that the background variables $\rho,u$ have their values $\rho(x,t),u(x,t)$ at the
location $x(t)$ of the soliton at the moment $t$. Then Eq.~(\ref{eq20}) can be written as one
of the Hamilton equations
\begin{equation}\label{eq34}
  \frac{dx}{dt}=u-2\rho+\sqrt{\rho(\rho-u)-\kappa^2/4}=\frac{\prt H}{\prt p}.
\end{equation}
It is convenient to introduce the variable $\phi$ instead of $\kappa$,
\begin{equation}\label{eq35}
  \kappa=2\sqrt{\rho(\rho-u)}\sin\phi,
\end{equation}
so that
\begin{equation}\label{eq36}
  \frac{dx}{dt}=\frac{\prt H}{\prt p}=u-2\rho+\sqrt{\rho(\rho-u)}\cos\phi.
\end{equation}

Another important relation follows from preservation of the Poincar\'{e}-Cartan integral 
invariant by the dispersionless flow (\ref{eq5}) (see Refs.~\cite{ks-23a,kamch-24}). This
means that in the completely integrable equations, as the DNLS equation, the wave number $k$ 
of a short wavelength packet propagating along large-scale background waves is only a function 
of the local values of $\rho$ and $u$, $k=k(\rho,u)$, which satisfies in the DNLS case the
equations \cite{ks-24}
\begin{equation}\label{eq37}
  \begin{split}
& \frac{\prt k^2}{\prt\rho}=4\left[\sqrt{k^2+4\rho(\rho-u)}-(2\rho-u)\right],\\
& \frac{\prt k^2}{\prt u}=-2\left[\sqrt{k^2+4\rho(\rho-u)}-2\rho\right],
\end{split}
\end{equation}
with the solution
\begin{equation}\label{eq38}
  k^2=(q+2\rho-u)^2-4\rho(\rho-u),
\end{equation}
where $q$ is an integration constant determined by the value of $k$ at some fixed point in
the $(\rho,u)$-plane. Then, using the Stokes reasoning (see Eq.~(\ref{eq21}) and the text below it),
we find that the inverse half-width $\kappa$ is related with the local background variables by
the formula
\begin{equation}\label{eq39}
  \kappa^2=4\rho(\rho-u)-(q+2\rho-u)^2.
\end{equation}
Its substitution into Eq.~(\ref{eq34}) with account of Eq.~(\ref{eq36}) give the integral of motion
\begin{equation}\label{eq40}
  u/2-\rho+\sqrt{\rho(\rho-u)}\cos\phi=q/2,
\end{equation}
where we assumed for definiteness that $q+2\rho-u>0$. Differentiation of this identity along
soliton's path yields
\begin{equation}\nonumber
  \begin{split}
&\frac12\frac{du}{dt}-\frac{d\rho}{dt}-\sqrt{\rho(\rho-u)}\sin\phi\frac{d\phi}{dt}\\
&+\frac{1}{2\sqrt{\rho(\rho-u)}}\left[(2\rho-u)\frac{d\rho}{dt}-\rho\frac{du}{dt}\right]=0,
\end{split}
\end{equation}
where $d/dt=\prt/\prt t+V\prt/\prt x$. We eliminate $\rho_t$ and $u_t$ with help of Eqs.~(\ref{eq5})
and find after elementary transformations the formula
\begin{equation}\label{eq41}
  \frac{d\phi}{dt}=-\left(\sqrt{\rho(\rho-u)}\right)_x\sin\phi.
\end{equation}

Now we can turn to derivation of the Hamiltonian $H$ and the canonical momentum $p$ for soliton's
motion. We look for the momentum in the form
\begin{equation}\label{eq42}
  p=\rho(\rho-u)f(\phi),
\end{equation}
where $f(\phi)$ is unknown function. Then integration of Eq.~(\ref{eq36}) with respect to $p$ gives
\begin{equation}\label{eq43}
  H=(u-2\rho)p+[\rho(\rho-u)]^{3/2}\int\cos\phi\,f'd\phi.
\end{equation}
To find $f(\phi)$, we use another Hamilton equation
\begin{equation}\label{eq44}
  \frac{dp}{dt}=-\frac{\prt H}{\prt x}.
\end{equation}
Substitution of Eqs.~(\ref{eq42}) and (\ref{eq43}) with the use of Eq.~(\ref{eq41}) leads to the equation
\begin{equation}\label{eq45}
  f'\sin\phi=3\int^{\phi}\cos\phi\,f'd\phi,
\end{equation}
which can be easily integrated to give
\begin{equation}\label{eq46}
  f=C(\phi-\sin\phi\cos\phi),
\end{equation}
where $C$ is an integration constant which appearance reflects invariance of the Hamilton equations
with respect to multiplication of $p$ and $H$ by the same constant factor. We choose it equal to
$C=2$ and express $\phi$ in terms of $\dot{x}=dx/dt=V$ and the local variables $\rho=\rho(x,t)$,
$u=u(x,t)$ with help of Eq.~(\ref{eq36}). As a result, we obtain
\begin{equation}\label{eq47}
\begin{split}
  &p=2\rho(\rho-u)\arccos\frac{\dot{x}+2\rho-u}{\sqrt{\rho(\rho-u)}}\\
&-2(\dot{x}+2\rho-u)\sqrt{\rho(\rho-u)-(\dot{x}+2\rho-u)^2},
\end{split}
\end{equation}
and
\begin{equation}\label{eq48}
  H=(u-2\rho)p+\frac43\left[\rho(\rho-u)-(\dot{x}+2\rho-u)^2\right]^{3/2}.
\end{equation}
It is worth noticing that these expressions reduce to similar expressions for the NLS soliton 
by means of replacements $\rho(\rho-u)\to\rho_{NLS}$, $u-2\rho\to u_{NLS}$ (see \cite{kamch-24}).

For practical applications, it is convenient to derive the Newton equation for the
soliton's motion.

\section{Newton equation}\label{sec5}

In Hamiltonian mechanics, $\dot{x}=\dot{x}(x,p)$, but it is impossible to solve explicitly Eq.~(\ref{eq47})
with respect to $\dot{x}$. Therefore, it is more convenient to transform the Hamilton equation
(\ref{eq44}) to the Newton equation for the soliton's acceleration $\ddot{x}$. Differentiation
of Eq.~(\ref{eq47}) with respect to time $t$ gives after obvious simplifications
\begin{equation}\label{eq49}
  \begin{split}
&\frac{dp}{dt}=2\left[(2\rho-u)\frac{d\rho}{dt}-\rho\frac{du}{dt}\right]
\arccos\frac{\dot{x}+2\rho-u}{\sqrt{\rho(\rho-u)}}\\
&-4\left(\ddot{x}+2\frac{d\rho}{dt}-\frac{du}{dt}\right)\sqrt{\rho(\rho-u)-(\dot{x}+2\rho-u)^2}.
\end{split}
\end{equation}
The right-hand side of Eq.~(\ref{eq44}) reads
\begin{equation}\label{eq50}
  \begin{split}
-\left.\frac{\prt H}{\prt x}\right|_p=& -(u_x-2\rho_x)p
-\left.\frac{\prt H^{(0)}}{\prt\rho}\right|_{\dot{x}}\rho_x\\
&-\left.\frac{\prt H^{(0)}}{\prt u}\right|_{\dot{x}}u_x
-\frac{\prt H^{(0)}}{\prt\dot{x}}\left.\frac{\prt\dot{x}}{\prt x}\right|_p,
\end{split}
\end{equation}
where $H^{(0)}=\frac43\left[\rho(\rho-u)-(\dot{x}+2\rho-u)^2\right]^{3/2}$. The derivative
$\left.\frac{\prt\dot{x}}{\prt x}\right|_p$ is to be obtained by differentiation of Eq.~(\ref{eq47})
with respect to $x$ at constant $p$ and this gives
\begin{equation}\label{eq51}
\begin{split}
  \left.\frac{\prt\dot{x}}{\prt x}\right|_p=&\frac{(\rho^2-\rho u)_x}{2\sqrt{\rho(\rho-u)-(\dot{x}+2\rho-u)^2}}\\
&\times \arccos\frac{\dot{x}+2\rho-u}{\sqrt{\rho(\rho-u)}}+u_x-2\rho_x.
\end{split}
\end{equation}
The other derivatives are trivial. Substitution of all the derivatives into Eq.~(\ref{eq50})
followed by equating the result to Eq.~(\ref{eq49}) and elimination of $\rho_t$ and $u_t$
with the use of the dispersionless equations
\begin{equation}\label{eq52}
\begin{split}
    &\rho_t+\left[\rho\left(u-\frac32\rho\right)\right]_x=0,\\
 & u_t+uu_x-(\rho u)_x=-U_x,
\end{split}
\end{equation}
where $U(x)$ is the potential of external forces, yield after some simplifications the Newton equation
\begin{equation}\label{eq53}
\begin{split}
 & 2\ddot{x}=\frac{\rho U_x}{\sqrt{\rho(\rho-u)-(\dot{x}+2\rho-u)^2}}\\
&\times \arccos\frac{\dot{x}+2\rho-u}{\sqrt{\rho(\rho-u)}}+\frac{d}{dt}(2\rho-u)-U_x.
\end{split}
\end{equation}
If there is no external potential, then the equation
\begin{equation}\label{eq54}
  2\ddot{x}=\frac{d}{dt}(2\rho-u)
\end{equation}
gives at once the integral of motion
\begin{equation}\label{eq55}
  2\dot{x}-2\rho+u=\mathrm{const}.
\end{equation}
Actually, it follows easily from Eqs.~(\ref{eq36}) and (\ref{eq40}) written for the case 
of a uniform background without external forces. This equation is convenient for finding the
soliton's paths for motion of solitons along rarefaction waves and other large-scale waves
evolving without influence of external forces. Its generalization (\ref{eq53}) is derived
under the assumption that non-uniformity of the background produced by the external potential
is more essential than change of the soliton's energy caused by its deformation: in our
derivation of the Hamiltonian we supposed that the soliton keeps its form in slightly
non-uniform background.

\begin{figure}[t]
\begin{center}
	\includegraphics[width = 7cm,height = 5cm]{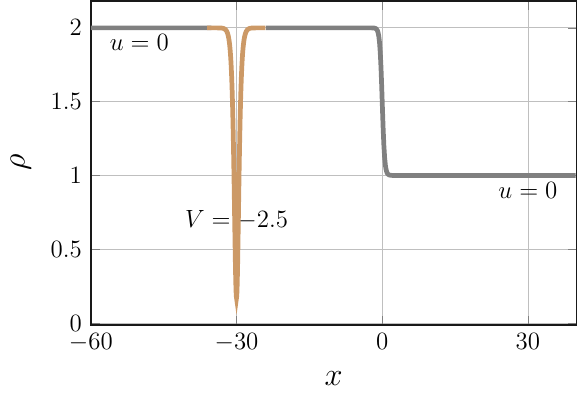}
\caption{Initial distribution of $\rho$ corresponding to interaction of a dark soliton with
a rarefaction wave; $\rho_L=2$, $\rho_R=1$, $l=30$, $V_0=-2.5$.
 }
\label{fig4}
\end{center}
\end{figure}

\begin{figure}[t]
\begin{center}
	\includegraphics[width = 7cm,height = 5cm]{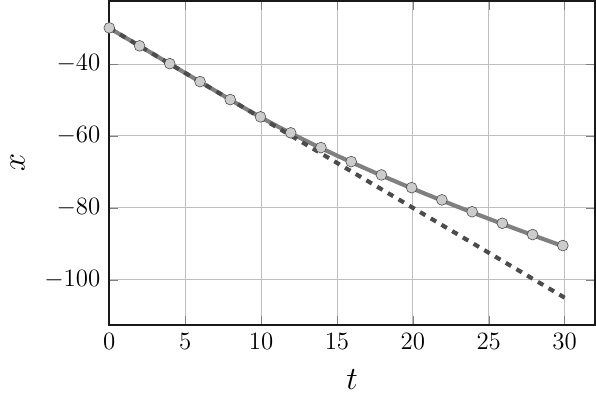}
\caption{Path of soliton along the rarefaction wave for initial parameters indicated
in Fig.~\ref{fig4}. Solid line corresponds to the analytical solution (\ref{eq59}),
dots to the numerical solution, and dashed line describes propagation of the soliton
with constant initial velocity $V_0$.
 }
\label{fig5}
\end{center}
\end{figure}

To check our analytical theory, we compared it with the results of exact numerical solutions 
of the DNLS equation (\ref{eq1}) for two typical physical situations. In the first one, the
soliton was propagating along a rarefaction wave produced by an initial discontinuity in the
distribution of $\rho$ (see Fig.~\ref{fig4}). In such a rarefaction wave the Riemann invariant
(see Eq.~(\ref{eq8})) $r_+=0$ remains constant and $r_-=-2\rho$, so that $v_-=-3\rho$. The 
self-similar solution of Eqs.~(\ref{eq7}) has the form $v_-=x/t$ and this gives the distribution
\begin{equation}\label{eq56}
  \rho(x,t)=
\left\{
\begin{array}{ll}
\rho_L,&\quad x<-3\rho_Lt,\\
-\frac{x}{3t},&\quad -3\rho_Lt<x<-3\rho_Rt,\\
\rho_R,&\quad x>-3\rho_Rt,
\end{array}
\right.
\end{equation}
We assume that the left edge of the rarefaction wave propagates to the left with velocity
$v_L=-3\rho_L$ faster, than the initial soliton velocity $V_0$. If the initial distance
between the soliton and the discontinuity equal to $l$, then after time $t_1=l/|v_L-V_0|$
the soliton starts its propagation along the rarefaction wave, so its motion is governed
by Eq.~(\ref{eq55}),
\begin{equation}\label{eq57}
  \frac{dx}{dt}+\frac{x}{3(t+t_1)}=V_0-\rho_L,
\end{equation}
whose solution must satisfy the initial condition
\begin{equation}\label{eq58}
  x(0)=-l+V_0t_1.
\end{equation}
This equation can be easily solved and the solution reads
\begin{equation}\label{eq59}
  x(t)=\frac34(V_0-\rho_L)(t+t_1)-\frac{3l}{4}\frac{t_1^{2/3}}{(t+t_1)^{1/3}}.
\end{equation}
As one can see, interaction of the soliton with the rarefaction wave leads to quite a considerable
effect and our analytical theory agrees very well with the numerical solution of the DNLS
equation.

\begin{figure}[t]
\begin{center}
	\includegraphics[width = 7cm,height = 5cm]{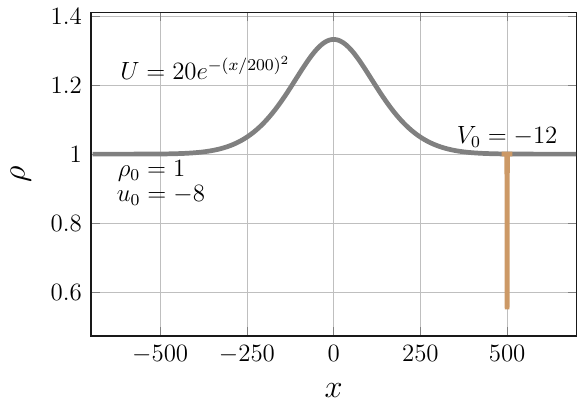}
\caption{Initial distribution of $\rho$ corresponding to interaction of a dark soliton with
an obstacle created by the potential (\ref{eq60}). 
 }
\label{fig6}
\end{center}
\end{figure}

\begin{figure}[t]
\begin{center}
	\includegraphics[width = 7cm,height = 5cm]{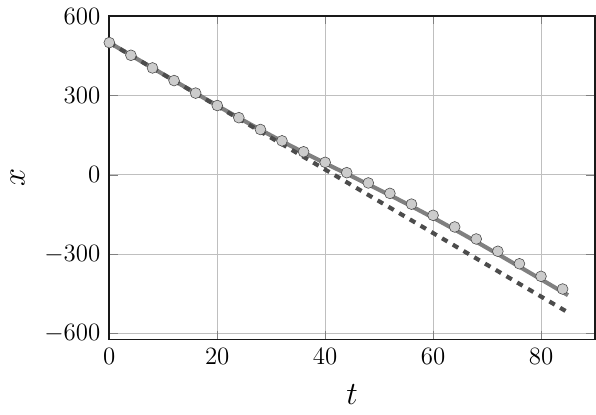}
\caption{Path of a soliton moving past an obstacle (\ref{eq60}). Solid line corresponds 
to the solution of Eq.~(\ref{eq53}),
dots to the numerical solution, and dashed line describes propagation of the soliton
with constant initial velocity $V_0$.
 }
\label{fig7}
\end{center}
\end{figure}

As another physical situation, we consider flow of the background past an obstacle created by the
potential
\begin{equation}\label{eq60}
  U(x)=20\exp\left[-\left(\frac{x}{200}\right)^2\right].
\end{equation}
The flow is stationary, so $\rho=\rho(x)$, $u=u(x)$ do not depend on time $t$ and they should be
obtained by solving the stationary dispersionless Eqs.~(\ref{eq52}). An easy calculation
(see, e.g., analogous problem for the NLS equation in Refs.~\cite{legk-09,ik-22}) yields the
solution in implicit form
\begin{equation}\label{eq61}
  F(u(x))=U(x),
\end{equation}
that is $u(x)=F^{-1}(U(x))$, where
\begin{equation}\label{eq62}
  F(u)=\frac13u\sqrt{9\rho_0^2-6\rho_0u_0+u^2}-\frac16u^2+\frac12u_0^2-\rho_0u_0.
\end{equation}
The Newton equation
\begin{equation}\label{eq63}
\begin{split}
 & 2\ddot{x}=\frac{\rho U_x}{\sqrt{\rho(\rho-u)-(\dot{x}+2\rho-u)^2}}\\
&\times \arccos\frac{\dot{x}+2\rho-u}{\sqrt{\rho(\rho-u)}}+2\rho_x-u_x-U_x
\end{split}
\end{equation}
should be solved with the initial conditions
\begin{equation}\label{eq64}
  x(0)=l,\quad \dot{x}=V_0,
\end{equation}
and the result is shown in Fig.~\ref{eq7} by a solid line, whereas dots correspond to the
numerical solution of Eq.~(\ref{eq1}) with the same initial condition in the form of the
soliton (\ref{eq33}) on the stationary background flow. Dashed line corresponds to the soliton
moving with constant initial velocity $V_0$. Again the analytical theory agrees very well 
with the numerical solution.

\section{Conclusion}

The problem of soliton's motion along non-uniform and time-dependent background, especially
in presence of external forces, is very difficult because of back reaction of the
counterflow caused by a moving soliton in the large-scale background wave. Previously,
different versions of the perturbation theory were developed for solving this problem,
but they were quite complicated and therefore applied to a very limited number of equations
(mainly, KdV and NLS equations). We suggested in Ref.~\cite{ks-23a} a new approach.
Although it is basically also perturbative, many difficulties are removed by imposing
the condition that the equations of soliton's motion must be Hamiltonian. Combining
this condition with Stokes' remark that some relationships for harmonic linear wave
can be converted into the relationships for solitons by replacement of the packet's wave
number $k$ by $i\kappa$, $\kappa$ being the soliton's inverse half-width, we reduce
derivation of the Hamiltonian and the canonical momentum of the soliton's motion to a
straightforward calculation. The resulting Hamilton equations can be transformed to
the Newton equation where the role of the external potential can be taken into account 
by its inclusion into the equations of the dispersionless (hydrodynamic) flow. The
effectiveness of this method was demonstrated in Ref.~\cite{kamch-24} for the NLS dark
soliton, where the results of Ref.~\cite{ik-22} were easily reproduced. In the present 
paper, we applied this method to the DNLS soliton which has quite a nontrivial dynamics
without reflection symmetry. Validity of our approach is confirmed by its good agreement 
with exact numerical solutions of the DNLS equation. We believe that this approach can
find many other applications.

\section*{Acknowledgments}

This research is funded by the research project FFUU-2024-0003 of the Institute of Spectroscopy
of the Russian Academy of Sciences (Sections~1--3) and by the RSF grant number~19-72-30028
(Section~4--6).

\end{document}